\documentstyle[12pt,aasms4]{article}
\makeatletter
\def\lnyoro{\mathrel{\mathpalette\gl@align<}}
\def\gnyoro{\mathrel{\mathpalette\gl@align>}}
\def\gl@align#1#2{\lower.6ex\vbox{\baselineskip\z@skip
\lineskip\z@\ialign{$\m@th
#1\hfil##\hfil$\crcr#2\crcr\sim\crcr}}}
\makeatother

\begin{document}

%\title{\bf ORIGIN OF COLOR GRADIENTS IN ELLIPTICAL GALAXIES IN THE
%HUBBLE DEEP FIELD NORTH} 
\title{\bf ORIGIN OF COLOR GRADIENTS IN ELLIPTICAL GALAXIES} 

\author{\bf Naoyuki Tamura$^{1}$, Chiaki Kobayashi$^{2}$, Nobuo
Arimoto$^{3}$, Tadayuki Kodama$^{2,4}$, \& Kouji Ohta$^{1}$}
\affil
{$^1$Department of Astronomy, Faculty of Science, Kyoto
University, Kyoto 606-8502, Japan}
\affil
{$^2$Department of Astronomy, School of Science, University of Tokyo, 
Tokyo 113-0033, Japan}
\affil
{$^3$Institute of Astronomy, School of Science, University of Tokyo,
Mitaka, Tokyo 181-8588, Japan}
\affil
{$^4$Physics Department, University of Durham, 
South Road, Durham, DH1 3LE, England}
\centerline{Aug. 3, 1999}
\centerline{email: tamura@kusastro.kyoto-u.ac.jp}
\authoremail{tamura@kusastro.kyoto-u.ac.jp}
\begin{abstract}

The origin of color gradients in elliptical galaxies is examined by
comparing model gradients with those observed in the Hubble Deep
Field. The models are constructed so as to reproduce color gradients in
local elliptical galaxies either by a metallicity gradient or by an age
gradient. By looking back a sequence of color gradient as a function of
redshift, the age - metallicity degeneracy is solved. The observed color
gradients in elliptical galaxies at $z=0.1-1.0$ agree excellently with
those predicted by the metallicity gradient, while they deviates
significantly from those predicted by the age gradient even at $z \sim
0.3$ and the deviation is getting larger with increasing redshift. This
result does not depend on cosmological parameters and parameters for an
evolutionary model of galaxy within a reasonable range.  Thus our
results clearly indicate that the origin of color gradients is not the
age but the stellar metallicity.

\keywords{galaxies: elliptical and lenticular, cD--- galaxies:
evolution--- galaxies: formation}

\end{abstract}

\section{INTRODUCTION}

Stellar populations in an elliptical galaxy are not uniform.  Stars at a
galaxy center are redder than those in the outer region and colors in a
galaxy are progressively getting bluer with an increasing radius (e.g.,
Vader et al. 1988; Franx, Illingworth, \& Heckman 1989; Peletier et
al. 1990a; Peletier, Valentijn, \& Jameson 1990b).  Peletier et al.
(1990a) made surface photometry in the $U$-, $B$-, and $R$-bands for a
sample of 39 nearby elliptical galaxies and examined their color
gradients in $U-R$ and $B-R$.  They found that typical color gradients
$\Delta(U-R)/\Delta \log r$ and $\Delta(B-R)/\Delta \log r$ are $-0.20$
mag$/$dex and $-0.09$ mag$/$dex, respectively, and demonstrated that a
dispersion of the color gradients is small, i.e., only 0.02 mag $/$dex
in both colors.  Since many of elliptical galaxies show radial gradients
in line strengths such as Mg$_{2}$, Fe$_{1}$(5270 \AA) and Fe$_{2}$(5335
\AA) (e.g., Carollo, Danziger, \& Buson 1993; Davies, Sadler, \&
Peletier 1993; Gonzalez 1993; Kobayashi \& Arimoto 1999), the color
gradients have been naively assumed to originate from a metallicity
gradient inside a galaxy.

However, such an interpretation for the origin of the color gradient is
premature, because stellar populations of either higher metallicity or
older age can make a galaxy redder. This problem, which is called the
{\it age-metallicity degeneracy}, was first pointed out by Worthey,
Trager, \& Faber (1996) and then discussed by Arimoto (1996). For
example, the degeneracy makes it difficult to interpret the origin of
the tight correlation between colors and magnitudes of elliptical
galaxies; brighter elliptical galaxies tend to have redder colors. This
correlation called color-magnitude (CM) relation can be excellently
reproduced by a metallicity sequence with a galactic wind model based on
a monolithic collapse scenario (e.g., Arimoto \& Yoshii 1987), where
more massive elliptical galaxies should be enriched more in metals and
thus become redder. However, Worthey et al. (1996) claimed that an age
sequence of elliptical galaxies can equivalently reproduce the CM
relation if brighter elliptical galaxies are older and thus redder. To
break this degeneracy, Kodama \& Arimoto (1997) built up two model
sequences (metallicity and age sequences) which are normalized to
reproduce the CM relation of ellipticals in Coma cluster by using their
evolutionary synthesis model. They compared the evolution of the model
CM relation with the observed relations of ellipticals in distant
clusters. The CM relations produced by a metallicity sequence agree with
the observed relations out to $z \sim 1$, while those produced by an age
sequence deviates from the observed ones significantly even at $z \sim
0.2-0.3$, showing the origin of the CM relation is primarily a
metallicity variation with a galaxy mass.  The CM relation can also be
reproduced in a hierarchical galaxy formation scenario (Kauffmann \&
Charlot 1998).  Although a model of Kauffmann \& Charlot (1998) allows a
more extended period of star formation in elliptical galaxies, it shows
that the CM relation is produced by a metallicity variation.  It is thus
worth emphasizing that the interpretation of CM relation with a
metallicity sequence is robust and independent of detailed assumptions
on galaxy formation processes.

Recently, the origin of color and line strength gradients in elliptical
galaxies has been discussed with elaborate models.  Martinelli,
Matteucci, \& Colafrancesco (1998) tried to reproduce the color
gradients by assuming that a galactic wind blows later in the inner part
of a galaxy due to a deeper potential well defined mainly by dark
matter.  Adopting a multi-zone model that takes into account gas
dynamics, local star formation, and chemical evolution, Tantalo et al.
(1998) reproduced radial gradients of colors and line strengths to some
extent.  Nevertheless, these detailed modelings are not fully
successful; the inner part of a galaxy becomes too iron enriched due to
an extended period of star formation there.

In this paper, adopting a much simpler approach without entering into
details of physical processes of galaxy formation and evolution, we try
to depict essential aspect of the origin of color gradients. Our
approach is similar to that adopted by Kodama \& Arimoto (1997) for
studying the origin of the CM relation. By using a population synthesis
model, we first make two different model galaxies, both of which can
reproduce a typical color gradient of elliptical galaxies at $z=0$, by
changing either mean stellar metallicity or age, and let evolve them
back in time. The evolution of color gradients thus predicted are then
confronted with the observed ones in distant ellipticals extracted from
the {\it Hubble Deep Field North} (HDF-N; Williams et al. 1996).  As was
demonstrated by Kodama \& Arimoto (1997), the best way to disentangle
the age and metallicity effects on galaxy colors is to look back
galaxies at high redshift.

It should be noted that dust extinction in elliptical galaxies may have
some effects on the color gradients (Goudfrooij \& de Jong 1994; Wise \&
Silva 1996). Witt, Thronson, \& Capuano (1992) calculated a radiative
transfer within elliptical galaxies by assuming a diffuse distribution
of dust and suggested that surface brightness profiles and color
gradients could be well reproduced by dust effects. In fact, it has been
believed that the IRAS detected FIR emission for about half of the
elliptical galaxies observed. However, since many of the detections were
made with a $\sim 3 \sigma$ threshold, a significant fraction of the
previously claimed detections may be spurious. Only $12-17\%$ of the
observed elliptical galaxies are detected with a sufficiently high
confidence level (Bregman et al. 1998). Therefore, in this paper, we
have chosen to focus on age and metallicity effects only. Effects of
dust extinction on the color gradients are still open and will be
studied in our subsequent paper.

This paper is organized as follows. The model prescription is given in
\S~2.  The sample selection and data reduction of elliptical galaxies in
the HDF-N are described in \S~3. The observed color gradients are
compared with the models in \S~4.  Discussion and conclusions are
given in \S~5 and \S~6, respectively.  The cosmological parameters
adopted throughout this paper are $H_{0} = 50$ km s$^{-1}$ Mpc$^{-1}$,
$\Omega_{0}=0.2$ and $\Lambda = 0$ unless otherwise noted.

\section {MODELS}

The color gradients can be reproduced by either a metallicity gradient
or an age gradient, but these effects are degenerate at $z=0$. We
attempt to break up such degeneracy by 
 confronting model color gradients with the observed ones
of ellipticals at high redshifts in the HDF-N.  We have built
two sequences of evolutionary models under the alternative assumptions:
1) the color gradients originate from the metallicity gradient of old
stellar populations, or 2) the color gradients arise from the age
gradient of stars which have the same metallicity within the galaxy.

The metallicity sequence is constructed as follows: We assume that the
star formation lasted progressively longer towards galaxy center; i.e.,
the galactic wind blew later in the inner region, so that the mean
stellar metallicity becomes higher and the colors become redder.  Even
at the galaxy center, however, the star formation stopped at early times
($t_{\rm gw}=0.83$ Gyr) and the stellar ages are almost the same at
everywhere within a galaxy; $\sim$ 15 Gyr at $z=0$ (see Tables 1 and 2).
The assumption seems to be justified, because the duration of star
forming activity is determined from the local dynamical potential
(Larson 1974) and should be longer in the inner region.  To build a
metallicity gradient, however, this hypothesis is not unique; other
hypotheses such as higher star formation efficiency, a larger yield
(i.e., a flatter initial mass function (IMF) slope), or higher
metallicity of infalling gas towards the galaxy center can also produce
the metallicity gradient. However, such details are not essential in the
present study, because so far as the mean stellar metallicity increases
in the inner part of a galaxy while stars are uniformly old, the
resulting evolution of color gradient is essentially the same. We should
stress this, because this paper aims for showing how color gradients of
stellar populations behave as a function of lookback time, not for
seeking the best model that can explain the observed color gradients of
sampled galaxies in the HDF-N. The galactic wind is conventionally
introduced to build up the metallicity gradient and it might be possible
that such gradients can also be established if elliptical galaxies
formed via hierarchical clustering.

The age sequence is constructed as follows: The star formation started
earlier in the inner regions, but lasted equally long ($t_{\rm gw}=0.83$
Gyr) everywhere in a galaxy. The period chosen here is the same as that
assumed for the center of metallicity sequence model.  In this way, the
mean stellar ages $t_{\rm age}$ become older and colors redder in the
inner regions, while the mean stellar metallicities are almost the same
everywhere (see Tables 1 and 2). It is hard to comprehend the physical
process to produce such age gradient, although it may occur if a series
of young dwarf galaxies accrete onto a massive galaxy and are tidally
disrupted at the outer region of a galaxy before falling into the
center.  However, it is not our aim anyway to construct a
physically motivated age sequence model.

Although we do not discuss in this paper in detail, we have tried to
build an alternative age sequence model in which star formation began
everywhere 15 Gyr ago and lasted longer at the outer parts of the
galaxy, keeping the same metallicity everywhere as that at the galaxy
center. As a result, the mean stellar ages become progressively younger
from the galaxy center towards the outer region. However, this model
fails to reproduce the observed color gradients in nearby ellipticals.
If old stellar populations are contained at the outer regions and if the
mean metallicity is as rich as that at the center, there is no way by
making their mean ages young to build up the color gradients as steep as
the observed ones.

Both metallicity and age sequences are constructed to reproduce typical
color gradients observed for nearby elliptical galaxies; $\Delta
(B-R)/\Delta \log r = - 0.09 \pm 0.02$ mag/dex (i.e., $-0.07$ and
$-0.11$ mag/dex)
%within a half of effective radius ($r_{\rm e}$)
and $B-R=1.633$ mag at $r=r_{\rm e}/10$ ($r_{\rm e}$ refers to an
effective radius) which is derived from the mean $B-R$ color at $r_{\rm
e}/2$ and $\Delta (B-R)/\Delta \log r = - 0.09$ mag/dex (Peletier et al. 
1990a).  We also construct the models which reproduce the color
gradients of $-0.09 \pm 0.04$ mag/dex (i.e., $-0.05$ and $-0.13$
mag/dex) within which the gradients of almost all the sample ellipticals
in Peletier et al. (1990a) are included.  Thus we have a set of four
model sequences each for the age sequence and the metallicity sequence.

To calculate gradients of photo-chemical properties of a galaxy, we
construct a galaxy model consisting of a 'shell' at each radius, and
assume that each 'shell' evolves independently.  The star formation rate
(SFR) in a 'shell' is proportional to the gas fraction (Schmidt law)
with a time scale of 0.1 Gyr and a primordial gas is supplied to the
shell with a rate of $\exp(-t/0.1\, {\rm Gyr})$.  An infall of a
primordial gas onto the shell may sound rather strange, since it is more
likely that the inner shells suffer from the infall of enriched gas from
the outer shells. However, a proper modeling of such effect is beyond
our scope. Since the infall rate we employed in the present study is
significantly high, the resulting evolutionary behavior of the color
gradient remains almost the same even if we adopt the 'simple' model
prescription instead of the infall model. For an IMF, a power-law mass
spectrum with a slope of $x=1.10$ is assumed in the range of
$0.05M_\odot \leq M \leq 50M_\odot$.  We note that this IMF slope is the
same as in Kodama et al. (1998), and was introduced to increase chemical
yields to reproduce the reddest end of the Coma CM relation.  The
nucleosynthesis yields of SNe Ia and SNe II are taken from Tsujimoto et
al.  (1995). The chemical enrichment by SNe Ia is calculated with a
metallicity dependent SN Ia rate, of which detail formulation is given
by Kobayashi, Tsujimoto \& Nomoto (1999). The spectral evolution is
calculated using a spectral synthesis database of Kodama \& Arimoto
(1997). The adopted parameter values for galactic wind epoch $t_{\rm
gw}$ and stellar age $t_{\rm age}$ at each radius for the four color
gradients are listed in Table 1.  The resulting photo-chemical
properties of a galaxy at $z=0$ are summarized in Table 2, including gas
abundances and colors at each radius.  It should be noted that our
models have an SN II-like abundance pattern, though the value [Mg/Fe]
$\sim +0.4$ is slightly larger than that suggested by the observational
estimates (Worthey, Faber \& Gonzalez 1992; Kobayashi \& Arimoto 1999);
a larger [Mg/Fe] ratio is a result of the assumed flat IMF ($x=1.10$).

\section {GALAXIES FOR COMPARISON}
\subsection {\it Sample Elliptical Galaxies}

To compare theoretical color gradients with those of elliptical galaxies
at high redshifts, the archival data of the HDF-N (Williams et al. 1996)
are used. Our sample galaxies are selected from those brighter than
$I_{814,AB} = 22$ mag, in such a way that we can derive reliable surface
brightness profiles and color gradients. All these galaxies have
spectroscopic redshifts. Elliptical galaxies are identified by using a
bulge-to-total ($B/T$) luminosity ratio derived by Marleau \& Simard
(1998) who obtained the $B/T$ ratios in the $I_{814}$-band of HDF
galaxies brighter than $I_{814}= 26$ mag by decomposing quantitatively
the surface brightness profile into the bulge and the disk
components. We define galaxies with $B/T > 0.5$ as ellipticals according
to Marleau \& Simard (1998). Our resulting sample consists of ten
elliptical galaxies with redshifts spanning from $z = 0.089$ to $1.015$
as listed in Table 3. It is important to note here that our sample
selection does not rely on any color information. Franceschini et
al. (1998) also made a sample of elliptical galaxies in the HDF using
surface brightness profiles. Their selection, however, is not based on
the bulge - disk decomposition.  Consequently, within $I_{814,AB} < 22$
mag, our sample is smaller than that of Franceschini et al. (1998),
except for the galaxy 4-241.1 at $z=0.318$.  The Franceschini et al's
sample includes several galaxies having the $B/T$ ratio smaller than
0.5, which are presumably disk dominant galaxies and thus are not
included in our sample (see also Kodama, Bower \& Bell 1999). It should
be noted that $B/T>0.5$ in the $I_{814}$-band may be slightly loose to
isolate elliptical galaxies.  $B/T>0.5$ in the $B$-band roughly
corresponds to local early-type galaxies (e.g., Simien \& de Vaucouleurs
1986).  By using bulge and disk models of Kodama et al. (1999), we found
that the $I_{814}$-band $B/T$ ratios of ellipticals are larger than
$\sim$ 0.7 at $z=0$ and change little from $z = 0$ to 1 in the
observer's frame.
%Early type galaxies with $B/T$ ratios of 0.5 in $B$-band at $z=0$
%(e.g., Simien \& de Vaucouleurs 1986) would have ratios of 0.7 in
%observed $I_{814}$-band when they are at $z \sim 1$, because of
%K-correction and an evolutionary effect which is modeled in 
%Kodama et al. (1999).
Thus our sample may include galaxies as late as Sa, though they must be
minority because most of the $B/T$ ratios of our sample galaxies are
larger than 0.7 (Table 3).

We measured $V_{606} - I_{814}$ colors of our sample galaxies and
present them in Table 3 and Figure 1. These colors are obtained for a 10
kpc aperture.  Seven of the sample galaxies have red colors consistent
with those for ellipticals with passive evolution as shown in Figure 1
(hereafter we call these galaxies `red ellipticals'), and the other
three galaxies (4-241.1, 2-251.0, and 4-928.0) have blue colors (see
also Figure 2; hereafter referred to as `blue ellipticals' for
convenience).  Azimuthally averaged surface brightness profiles of the
sample ellipticals in the $I_{814}$-band are shown in Figure 2.  The
average is taken along an ellipse fitted to an isophote of the
$I_{814}$-band image.  These profiles are well represented by an
$r^{1/4}$ law for both the red ellipticals and the blue ellipticals.
Effective radii ($r_{\rm e}$) of the sample ellipticals are obtained by
the $r^{1/4}$ fitting and results are shown in Table 3.  The fitting is
done by removing data points in the inner and outer regions, since they
are unreliable due to an effect of a point spread function (PSF) and low
signal-to-noise ratios (S/Ns), respectively. The obtained effective
radii range from $r_e=$1.4 to 7.9 kpc, which are typical for nearby
giant ellipticals (e.g., Bender, Burstein, \& Faber 1992).

\subsection {\it Color Gradients}

To examine color gradients of our sample galaxies, the images in F606W
($V_{606}$)- and F814W ($I_{814}$)-bands were used, because S/Ns of the
images are better in these bands than in the other bands (F300W and
F450W).  First, the sky value around each object was determined by using
the "phot" task in the IRAF apphot package ("mode" in an annulus with an
inner radius of mostly $3^{\prime\prime} \sim 4^{\prime\prime}$ and a
width of 0.$^{\prime\prime}4$ is adopted) and was subtracted.  Next, the
angular resolutions of the blue images and the red images were adjusted
and a $V_{606} - I_{814}$ color map of each galaxy was made.  Figure 3
shows the resulting color maps.  Most of the color maps for the red
ellipticals show symmetric structure and ordinary color gradient.  In
two cases (2-456.0 and 2-121.0), slight asymmetry in the color
distribution is seen.  This does not seem to be caused by a misalignment
of the images.  For the blue ellipticals, variety of the color
distribution is seen, which we will discuss in \S 5.1.
Next, a $V_{606} -
I_{814}$ color profile is derived with the azimuthally averaged radial
profiles as a function of semi-major radius from the galaxy center.
Resulting color profiles are shown in Figures 4 and 5 for the red
ellipticals and in Figure 7 for the blue ellipticals.
It should be noted that in plotting observed color profiles, 
a zero-point of the observed color for each galaxy is shifted to
be compared with those by the models.
The amounts of these shifts are given in the last column of Table 3.
Most of the shifts are less than 0.2 mag; for intrinsically luminous
objects the colors tend to be shifted to bluer colors and vice versa.
Thus the color differences are presumably caused by an effect of the
color-magnitude relation of elliptical galaxies.
The zero-point offsets could also come from
photometric errors in HST data (e.g., Holtzman et al. 1995; Ellis et
al. 1997; Kodama et al. 1998) and from uncertainty of the population
synthesis model.
Needless to say, zero-point offsets do not affect the color gradients.
An error bar attached to each data point in the color
profile includes an photometric error, a local sky subtraction error,
and a dispersion of colors along each elliptical isophote.  The data
points of 2-456.0 and 2-121.0 have relatively large error bars in the
middle region of the profiles; these objects have asymmetric color
distributions which results in the relatively large dispersions of
colors in the regions.

Finally, slopes of the color profiles
$\Delta(V_{606} - I_{814}) / \Delta\log (r/ r_{\rm e})$ of the sample
galaxies are derived by applying a least square fit to the color
profiles. This fit was done after removing the unreliable data points in
the outer region ($r > r_{\rm e}$) and in the innermost region.  The
removed data points in the inner most regions are shown by crosses in
the color profiles (Figures 4 and 5); they clearly deviate from the
$r^{1/4}$ fits presumably due to the effect of a PSF as seen in Figure
2.  The slopes of the color profiles for the blue ellipticals presented
in Figure 7 are not derived.

\section {ORIGIN OF COLOR GRADIENTS}

Once the spectra are calculated at various radii of a model galaxy in
such a way that they reproduce the color gradient $\Delta(B-R)/ \Delta
\log (r/r_{\rm e}) = -0.09$ mag/dex at $z=0$, the model gradient in any
colors at any redshifts can be predicted to compare with the observed
gradients. The model color gradients at the redshift of each sample
elliptical galaxy are over-plotted in each panel of Figures 4 and 5 for
the age gradient and the metallicity gradient models,
respectively. Solid and dotted lines in each panel show the gradients in
$V_{606} - I_{814}$ color in the observer's frame corresponding to the
gradients of $-0.09 \pm 0.02$ mag/dex and $-0.09 \pm 0.04$ mag/dex in
the $B-R$ color at $z = 0$, respectively.

It is clearly shown that the model gradient made by the age gradient
begins to deviate from the observed ones at a redshift of $z \sim 0.3$
and the deviation is getting worse as a redshift increases (Figure 4).
On the contrary, the model gradient made by the metallicity gradient
agrees well with the observed color gradients within the effective
radius at any redshift from $z \sim 0.1$ to $\sim 1$ (Figure 5).
Considering that the model gradients are calibrated only at $z=0$, we
insist that the agreement between the model and the observed gradients
is excellent.  Such agreement is more clearly seen in Figure 6 which
shows both the observed color gradients and the model gradients as a
function of redshift. Solid and dotted lines show the evolution of model
color gradients in $V_{606} - I_{814}$ in the observer's frame for the
metallicity gradient and the age gradient, respectively. These model
gradients are calculated in a region of $-1 \le \log r/r_{\rm e} \le 0$
for the metallicity gradient and for the age gradient at $z<0.5$, and $-1
\le \log r/r_{\rm e} \le - 0.5$ for the age gradient at $z>0.5$.  Dots
show the observed gradients of the sample galaxies within $r_{\rm e}$
calculated by a least square fitting to the data points shown as filled
circles in Figures 4 and 5. Error bars indicate the fitting errors of
1$\sigma$.

\section {DISCUSSION}

\subsection {\it Blue Galaxies in Our Sample}

Since our sample is selected independently of galaxy colors, relatively
blue elliptical-like galaxies are included in our sample (3/10). Among
these three, 4-241.1 ($z=0.318$) has a nearly flat color profile with a
slightly bluer color in the inner part (Figure 3 and the top panel of
Figure 7).  The spectrum of this galaxy (from the Hawaii Active Catalog;
Cowie 1997) shows Balmer absorption lines shortward of 4000\AA\ in the
rest frame as well as emission lines such as [OII]$\lambda$3727,
H$\beta$, [OIII]$\lambda$4959, $\lambda$ 5007, and H$\alpha$.  Thus this
galaxy is expected to have some young stellar populations. The galaxy
2-251.0 ($z = 0.960$) has a color gradient in the opposite sense to
those of ordinary ellipticals; i.e., the inner region of the galaxy is
bluer than the outer region (Figure 3 and the middle panel of Figure 7).
Since this galaxy has been suggested as an AGN (Franceschini et
al. 1998), this opposite gradient is probably caused by AGN and/or
associating starburst activity in the central part of the
galaxy. Although both galaxies seem to have the younger ages, our model
by the age gradient cannot reproduce their color gradients.
Accordingly, neither the metallicity gradient nor the age gradient can
explain their observed color gradients. These galaxies presumably have
on-going star formation (or post-starburst) in the inner part of the
galaxies, which may be caused by a galaxy interaction as expected from
their rather disturbed morphology.  Our model cannot apply to these
cases.  The other galaxy 4-928.0 ($z = 1.015$) has a rather steep color
gradient as seen in Figure 3 and the bottom panel of Figure 7.  The
observed color profile of the galaxy is shifted by 0.4 mag to be
compared with the model color gradients (solid and dotted lines). The
observed color gradient of this galaxy may be explained by the age
gradient if we tune ages of stellar populations in the model galaxy.
However, the mean age is set to be 0.2 Gyr at $r_{\rm e}$ and 9 Gyr at
the innermost point. Such a large age at the center cannot be realized
at this redshift ($z=1.015$) unless a different set of cosmological
parameters is adopted. Finally, note that colors of these galaxies are
not consistent with those of passively evolving ellipticals formed at
high $z$. Much larger sample is needed in future works to find whether
the population of the blue galaxies is really the minority in the whole
population of elliptical galaxies.

\subsection {\it Gradients of Metal Absorption Line Strengths}

Kobayashi \& Arimoto (1999) showed that central velocity dispersions of
elliptical galaxies correlate well with their mean metallicities. They
found, however, that this correlation has a significant scatter. They
also showed that Mg$_{2}$ gradients do not correlate with any physical
parameters including a galaxy mass.  Such a large dispersion and an
absence of the correlation are contrary to what monolithic collapse
simulations predicted. As the common origin for the large dispersion and
the absence of the correlation, they suggested the following
possibilities: (1) a significant age spread among elliptical galaxies,
(2) age gradients inside elliptical galaxies, (3) dust obscuration, (4)
effects of merging on metallicity gradients inside elliptical galaxies,
and (5) intrinsic scatter which elliptical galaxies have had since they
formed. Since we confirm in this paper that the color gradients are
caused by the metallicity gradient, we can reject the possibility (2).
%that the age gradient has a significant effect on gradients of metal
%line indices and make large dispersions of them.

There seems to be a problem that color gradients in elliptical galaxies
do not correlate with their Mg$_{2}$ gradients, though only a small
number of elliptical galaxies are examined both for color gradient and
Mg$_{2}$ gradient (Peletier 1989).  However, the absence of this
correlation may be a spurious result caused by a different way to derive
the gradients between the color and Mg$_{2}$ gradients. Some effects of
dust extinction and/or, possibly, observational errors on both gradients
might suppress the true correlation.  In any case, this problem will be
investigated in detail in our forthcoming paper.

Worthey et al. (1992) suggested that the color-magnitude relation of
elliptical galaxies may originate from the fact that more luminous
ellipticals tend to contain more Mg relative to Fe.  One can claim that
it could be possible that the same holds for the internal gradients.
However, Kobayashi \& Arimoto (1999) have recently carried out a careful
study of line strength gradients of 80 elliptical galaxies having the
best quality of spectra but find no evidence supporting for the [Mg/Fe]
gradient. Therefore, we conclude that the color gradients are not
originating from the gradient of the [Mg/Fe] ratio.

\section {CONCLUSIONS}

The origin of color gradients in elliptical galaxies is examined.  A
typical color gradient of nearby elliptical galaxies is reproduced by
two alternative model sequences, metallicity gradient and age gradient.
This age-metallicity degeneracy is broken by looking back the evolution
of color gradient towards high redshifts; the predicted color gradients
at high redshifts are compared with the observed color gradients of ten
elliptical galaxies in the HDF-N out to $z\sim1$. We find that the
observed color gradients of the seven red galaxies, whose colors are
consistent with those of passively evolving galaxies, are in excellent
agreement with the metallicity gradient at any redshift.
%The remaining three bluer galaxies in our sample have
%peculiar color gradients and cannot be reproduced in the same
%metallicity sequence model, but no galaxy except one can support the age
%origin of the color gradient in elliptical galaxies.
These conclusions are independent of cosmological parameters and
parameters for an evolutionary model of galaxy (IMF and SFR) within a
reasonable range.  Thus we conclude that the primary origin for the
color gradient in elliptical galaxies is the metallicity gradient in the
old stellar populations.

\acknowledgements

This work was financially supported in part by Grant-in-Aids for the
Scientific Research (No. 0940311 and 09740173) by the Japanese Ministry
of Education, Culture, Sports and Science. CK and TK thank Research
Fellowships of the Japan Society for the Promotion of Science (JSPS) for
Young Scientists.

\newpage

\begin{deluxetable}{lrrrrrrrrr}
\footnotesize
\tablewidth{0pt}
\tablecaption{Input Parameters \label{tab:input}}
\tablehead{
\colhead{$\log r/r_{\rm e}$} & 
\colhead{$-1$} & \colhead{$-0.875$} &
\colhead{$-0.75$} & \colhead{$-0.625$} &
\colhead{$-0.5$} & \colhead{$-0.375$} &
\colhead{$-0.25$} & \colhead{$-0.125$} & \colhead{$0$}
}
\startdata
\cutinhead{Metallicity Sequence ($t_{\rm age}=15$ Gyr)}

$\Delta (B-R)/\Delta \log r$ & & & & & $t_{\rm gw}$ [Gyr] & & & \nl
$-0.05$ & 0.83 & 0.74 & 0.64 & 0.58 & 0.53 & 0.49 & 0.45 & 0.42 & 0.39 \nl
$-0.07$ & 0.83 & 0.69 & 0.59 & 0.52 & 0.47 & 0.42 & 0.38 & 0.35 & 0.32 \nl 
$-0.11$ & 0.83 & 0.63 & 0.52 & 0.43 & 0.37 & 0.32 & 0.29 & 0.26 & 0.23 \nl
$-0.13$ & 0.83 & 0.60 & 0.48 & 0.39 & 0.33 & 0.29 & 0.26 & 0.22 & 0.20 \nl

\cutinhead{Age Sequence ($t_{\rm gw}=0.83$ Gyr)}

$\Delta (B-R)/ \Delta \log r$ & & & & & $t_{\rm age}$ [Gyr] & & & \nl
$-0.05$ & 15.00 & 14.35 & 13.65 & 13.00 &12.40 &11.20 &10.20 & 9.75 & 9.20 \nl
$-0.07$ & 15.00 & 14.00 & 13.10 & 12.20 &10.60 & 9.75 & 9.00 & 8.60 & 7.80 \nl
$-0.11$ & 15.00 & 13.60 & 12.00 &  9.95 & 8.90 & 7.85 & 7.18 & 6.70 & 6.20 \nl
$-0.13$ & 15.00 & 13.35 & 11.10 &  9.35 & 8.35 & 7.20 & 6.65 & 6.10 & 5.63 \nl
\enddata
\tablenotetext{}{Note.--- The second and third rows are the models
constructed for the typical color gradients of $\Delta(B-R)/\Delta\log
r=-0.09 \pm 0.02$ mag/dex.  The models in the first and fourth rows are
for $\Delta(B-R)/\Delta\log r=-0.09\pm0.04$ mag/dex, within which
gradients of almost all ellipticals in Peletier et al.(1990a) are
included.}
\end{deluxetable}

\begin{deluxetable}{llrrrrrrrrr}
\tablenum{2}
\footnotesize
\tablewidth{0pt}
\tablecaption{Properties of Model Galaxies at $z=0$}
%\tablenotemark{a} \label{tab:output}}
\tablehead{
\colhead{$\log r/r_{\rm e}$} & &
\colhead{$-1$} & \colhead{$-0.875$} &
\colhead{$-0.75$} & \colhead{$-0.625$} &
\colhead{$-0.5$} & \colhead{$-0.375$} &
\colhead{$-0.25$} & \colhead{$-0.125$} & \colhead{$0$}
}

\startdata
\cutinhead{Metallicity Sequence ($t_{\rm age}=15$ Gyr)}

%$t_{\rm age}$ 
% & & 15.00 & 15.00 & 15.00 & 15.00 & 15.00 & 15.00 & 15.00 & 15.00 & 15.00 \nl
$\log Z/Z_\odot$ 
 &  & 0.254 & 0.240 & 0.219 & 0.201 & 0.182 & 0.168 & 0.142 & 0.123 & 0.101 \nl
 &  & 0.254 & 0.230 & 0.204 & 0.177 & 0.153 & 0.123 & 0.092 & 0.066 & 0.034 \nl
 &  & 0.254 & 0.216 & 0.177 & 0.136 & 0.084 & 0.034 &-0.002 &-0.046 &-0.080 \nl
 &  & 0.254 & 0.210 & 0.158 & 0.101 & 0.056 &-0.002 &-0.046 &-0.118 &-0.140 \nl
[Fe/H] 
 & &-0.069 &-0.088 &-0.112 &-0.131 &-0.151 &-0.165 &-0.192 &-0.213 &-0.236 \nl
 & &-0.069 &-0.100 &-0.128 &-0.155 &-0.181 &-0.213 &-0.245 &-0.273 &-0.307 \nl
 & &-0.069 &-0.115 &-0.155 &-0.199 &-0.254 &-0.307 &-0.346 &-0.394 &-0.431 \nl
 & &-0.069 &-0.121 &-0.175 &-0.236 &-0.284 &-0.346 &-0.394 &-0.474 &-0.497 \nl
[Mg/Fe] 
 & & 0.467 & 0.473 & 0.475 & 0.475 & 0.476 & 0.477 & 0.478 & 0.479 & 0.481 \nl
 & & 0.467 & 0.474 & 0.475 & 0.476 & 0.478 & 0.479 & 0.481 & 0.483 & 0.486 \nl
 & & 0.467 & 0.475 & 0.476 & 0.479 & 0.482 & 0.486 & 0.489 & 0.493 & 0.496 \nl
 & & 0.467 & 0.475 & 0.477 & 0.481 & 0.484 & 0.489 & 0.493 & 0.500 & 0.503 \nl
$U-R$ 
 & & 2.166 & 2.158 & 2.144 & 2.132 & 2.118 & 2.109 & 2.089 & 2.076 & 2.060 \nl
 & & 2.166 & 2.152 & 2.134 & 2.115 & 2.098 & 2.076 & 2.055 & 2.036 & 2.014 \nl
 & & 2.166 & 2.143 & 2.115 & 2.085 & 2.049 & 2.014 & 1.988 & 1.957 & 1.933 \nl
 & & 2.166 & 2.139 & 2.101 & 2.060 & 2.029 & 1.988 & 1.957 & 1.904 & 1.888 \nl
$B-R$ 
 & & 1.632 & 1.627 & 1.620 & 1.614 & 1.608 & 1.604 & 1.595 & 1.589 & 1.583 \nl
 & & 1.632 & 1.624 & 1.615 & 1.606 & 1.599 & 1.589 & 1.580 & 1.572 & 1.563 \nl
 & & 1.632 & 1.619 & 1.606 & 1.593 & 1.578 & 1.563 & 1.552 & 1.539 & 1.528 \nl
 & & 1.632 & 1.617 & 1.600 & 1.583 & 1.569 & 1.552 & 1.539 & 1.516 & 1.509 \nl
$V_{606}-I_{814}$ 
 & & 0.576 & 0.574 & 0.572 & 0.569 & 0.567 & 0.565 & 0.562 & 0.559 & 0.556 \nl
 & & 0.576 & 0.573 & 0.570 & 0.566 & 0.563 & 0.559 & 0.555 & 0.552 & 0.548 \nl
 & & 0.576 & 0.571 & 0.566 & 0.561 & 0.554 & 0.548 & 0.543 & 0.536 & 0.531 \nl
 & & 0.576 & 0.570 & 0.564 & 0.556 & 0.550 & 0.543 & 0.536 & 0.525 & 0.522 \nl

\cutinhead{Age Sequence ($t_{\rm gw} = 0.83$ Gyr)}

$t_{\rm age}$ 
 & & 15.00 & 14.35 & 13.65 & 13.00 & 12.40 & 11.20 & 10.20 &  9.75 &  9.20 \nl
 & & 15.00 & 14.00 & 13.10 & 12.20 & 10.60 &  9.75 &  9.00 &  8.60 &  7.80 \nl
 & & 15.00 & 13.60 & 12.00 &  9.95 &  8.90 &  7.85 &  7.18 &  6.70 &  6.20 \nl
 & & 15.00 & 13.35 & 11.10 &  9.35 &  8.35 &  7.20 &  6.65 &  6.10 &  5.63 \nl
$\log Z/Z_\odot$ 
 & & 0.254 & 0.254 & 0.254 & 0.254 & 0.254 & 0.254 & 0.254 & 0.254 & 0.254 \nl
 & & 0.254 & 0.254 & 0.254 & 0.254 & 0.254 & 0.254 & 0.254 & 0.254 & 0.253 \nl
 & & 0.254 & 0.254 & 0.254 & 0.254 & 0.254 & 0.253 & 0.253 & 0.253 & 0.253 \nl
 & & 0.254 & 0.254 & 0.254 & 0.254 & 0.253 & 0.253 & 0.253 & 0.253 & 0.253 \nl
[Fe/H] 
 & &-0.069 &-0.069 &-0.069 &-0.070 &-0.070 &-0.070 &-0.070 &-0.070 &-0.070 \nl
 & &-0.069 &-0.069 &-0.070 &-0.070 &-0.070 &-0.070 &-0.070 &-0.070 &-0.070 \nl
 & &-0.069 &-0.069 &-0.070 &-0.070 &-0.070 &-0.070 &-0.070 &-0.070 &-0.070 \nl
 & &-0.069 &-0.070 &-0.070 &-0.070 &-0.070 &-0.070 &-0.070 &-0.070 &-0.070 \nl
[Mg/Fe] 
 & & 0.467 & 0.467 & 0.467 & 0.467 & 0.467 & 0.467 & 0.467 & 0.467 & 0.467 \nl
 & & 0.467 & 0.467 & 0.467 & 0.467 & 0.467 & 0.467 & 0.467 & 0.467 & 0.467 \nl
 & & 0.467 & 0.467 & 0.467 & 0.467 & 0.467 & 0.467 & 0.467 & 0.467 & 0.467 \nl
 & & 0.467 & 0.467 & 0.467 & 0.467 & 0.467 & 0.467 & 0.467 & 0.467 & 0.467 \nl
$U-R$ 
 & & 2.167 & 2.154 & 2.140 & 2.126 & 2.114 & 2.099 & 2.082 & 2.068 & 2.054 \nl
 & & 2.167 & 2.147 & 2.129 & 2.112 & 2.089 & 2.068 & 2.047 & 2.031 & 2.012 \nl
 & & 2.167 & 2.139 & 2.110 & 2.074 & 2.043 & 2.013 & 1.988 & 1.963 & 1.938 \nl
 & & 2.167 & 2.133 & 2.098 & 2.058 & 2.023 & 1.989 & 1.961 & 1.932 & 1.904 \nl
$B-R$ 
 & & 1.633 & 1.627 & 1.621 & 1.614 & 1.609 & 1.603 & 1.596 & 1.590 & 1.584 \nl
 & & 1.633 & 1.624 & 1.615 & 1.608 & 1.599 & 1.590 & 1.581 & 1.574 & 1.566 \nl
 & & 1.633 & 1.620 & 1.607 & 1.593 & 1.579 & 1.567 & 1.553 & 1.539 & 1.525 \nl
 & & 1.633 & 1.618 & 1.603 & 1.586 & 1.570 & 1.553 & 1.538 & 1.522 & 1.506 \nl
$V_{606}-I_{814}$ 
 & & 0.577 & 0.575 & 0.573 & 0.570 & 0.569 & 0.566 & 0.564 & 0.561 & 0.560 \nl
 & & 0.577 & 0.574 & 0.571 & 0.569 & 0.565 & 0.561 & 0.559 & 0.558 & 0.563 \nl
 & & 0.577 & 0.573 & 0.568 & 0.562 & 0.559 & 0.562 & 0.557 & 0.547 & 0.537 \nl
 & & 0.577 & 0.572 & 0.566 & 0.560 & 0.559 & 0.557 & 0.546 & 0.534 & 0.524 \nl
 
\enddata
\tablenotetext{}{Note.--- The first, second, third and forth rows of each
physical quantity correspond to those for the color gradient of $\Delta
(B-R)/\Delta \log r = -0.05$, $-0.07$, $-0.11$, and $-0.13$ mag/dex,
respectively.}
\end{deluxetable}

\begin{table}
\tablenum{3}

 \caption{Sample Elliptical Galaxies}

 \begin{tabular}{cccccccc} 
 \hline\hline
  ID\tablenotemark{a} & $z$\tablenotemark{b} &
  $I_{814}$\tablenotemark{c} & $M_{F450W,AB}$\tablenotemark{c} &
 ($V_{606} - I_{814}$)$_{AB}$ & $r_{\rm e}$ & $B/T$\tablenotemark{d} & 
  $\Delta(V_{606} - I_{814})_{AB}$\tablenotemark{e}\\ 
          &       & (mag) &   (mag)  & (mag) & (kpc) & & (mag) \\ \hline
  2-456.0\tablenotemark{f} & 0.089 & 18.2\tablenotemark{g} &
  $-$19.20\tablenotemark{h} & 0.610 & 1.4 & - & $-$0.02\\
  4-241.1 & 0.318 & 20.33 & $-$20.94 & 0.422 & 3.1 & 0.66 & - \\
  2-121.0 & 0.475 & 20.03 & $-$20.94 & 1.071 & 3.5 & 0.87 & 0.10 \\
  3-790.0 & 0.562 & 20.93 & $-$21.92 & 1.152 & 2.8 & 0.63 & 0.21 \\
  3-321.0 & 0.677 & 20.92 & $-$22.17 & 1.490 & 7.0 & 0.81 & 0.09 \\
  4-744.0 & 0.764 & 20.41 & $-$23.12 & 1.609 & 5.1 & 0.79 & 0.06 \\
  4-493.0 & 0.847 & 21.17 & $-$22.89 & 1.670 & 4.0 & 0.81 & 0.02 \\
  2-251.0 & 0.960 & 20.86 & $-$23.65 & 1.055 & 7.9 & 1.00 & - \\
  4-752.0 & 1.013 & 20.88 & $-$24.10 & 1.855 & 6.9 & 0.87 & $-$0.14\\
  4-928.0 & 1.015 & 21.75 & $-$23.03 & 1.347 & 3.7 & 0.81 & $-$0.4 \\ 
 \hline\hline
 \end{tabular}
\tablenotetext{a}{Williams et al. (1996).}
\tablenotetext{b}{Cohen et al. (1996).} 
\tablenotetext{c}{Abraham et al. (1999). We converted the $M_{F450W,AB}$
magnitudes into those in the adopted cosmology.}
\tablenotetext{d}{Bulge-to-total luminosity ratio in the $I_{814}$-band taken from Marleau \&
Simard (1998).}
\tablenotetext{e}{Shifted zero-points to compare the model color
gradients for typical elliptical galaxies with observed ones (see \S
4).}
\tablenotetext{f}{The $B/T$ ratio of this object is not in Marleau \&
Simard (1998). However, we add it to our sample because its spectrum
(from the Hawaii active catalog; Cowie 1997) as well as morphology 
show this galaxy is clearly a typical E or S0.}
\tablenotetext{g}{Bouwens, Broadhurst, \& Silk (1998).}  
\tablenotetext{h}{$b_{J}$-band absolute magnitude calculated by Bouwens
et al. (1998). We also converted the magnitude into that in the adopted
cosmology.}
\end{table}

\clearpage

\centerline{\bf Figure Caption}

\noindent
{Fig. 1 --- $V_{606} - I_{814}$ colors of our sample galaxies in a 10
kpc aperture as a function of redshift.  Solid line shows a color track
for a passively evolving elliptical and dotted line for a non-evolving
elliptical.  Filled circles indicate red galaxies whose colors are
nearly consistent with that of the passively evolving or non-evolving
ellipticals.  Open circles indicate the remaining bluer galaxies.}

\noindent
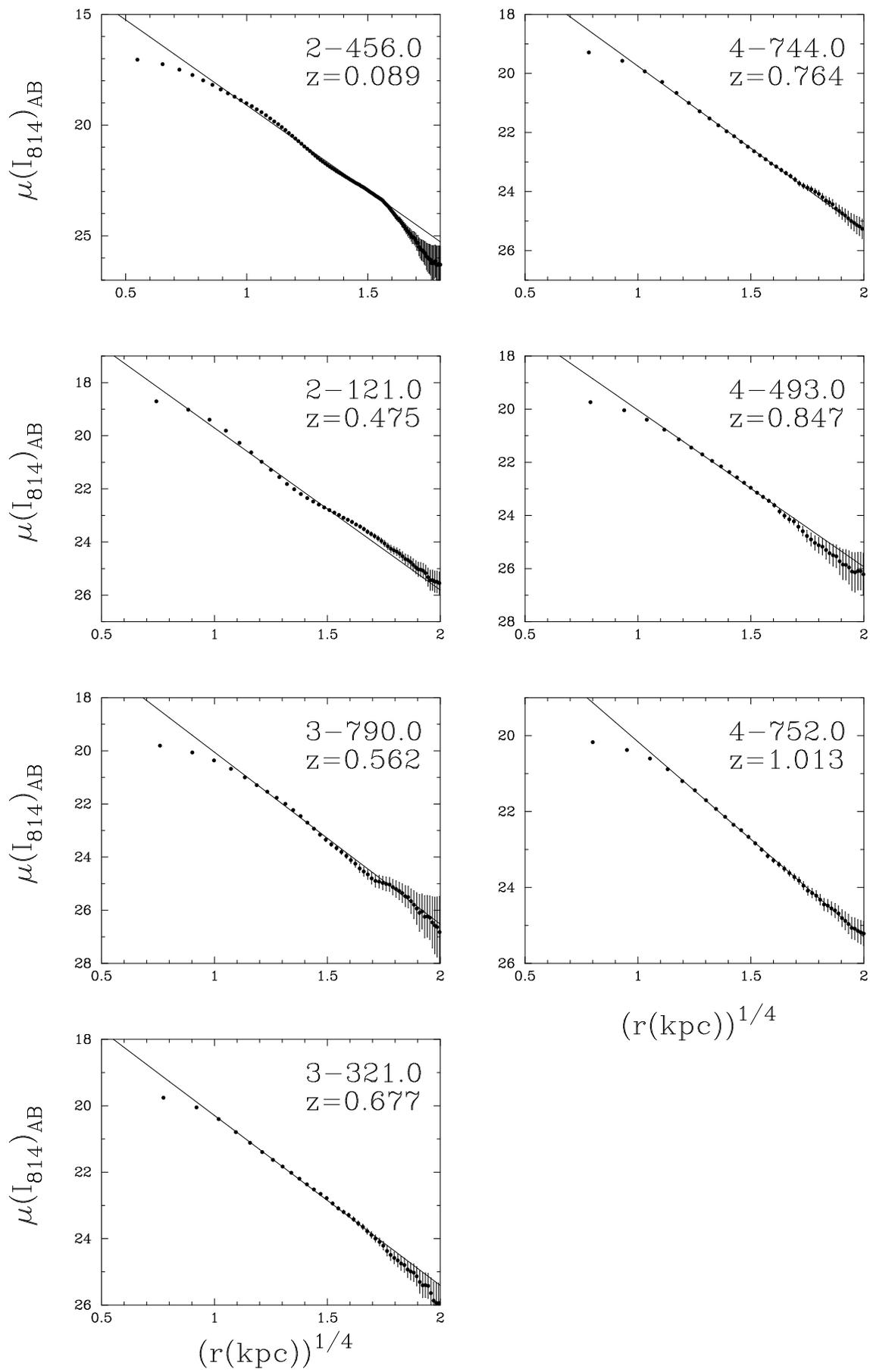
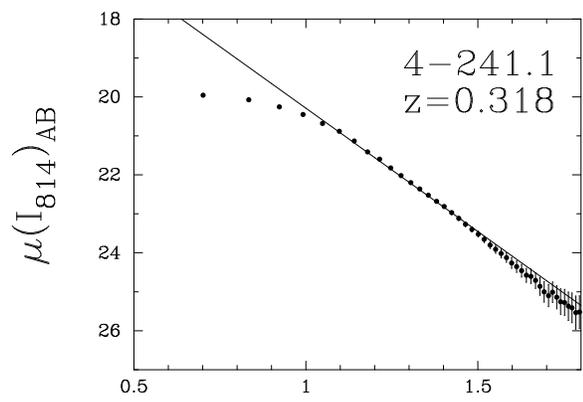
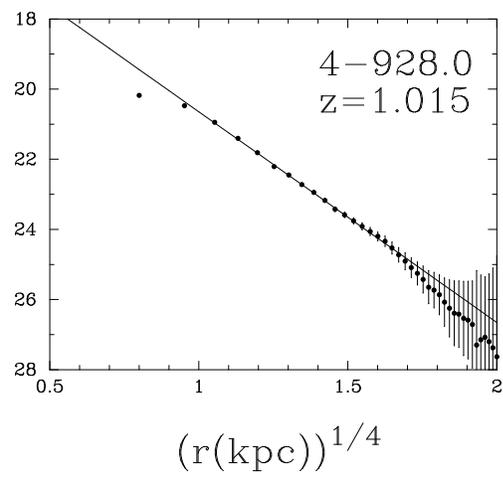
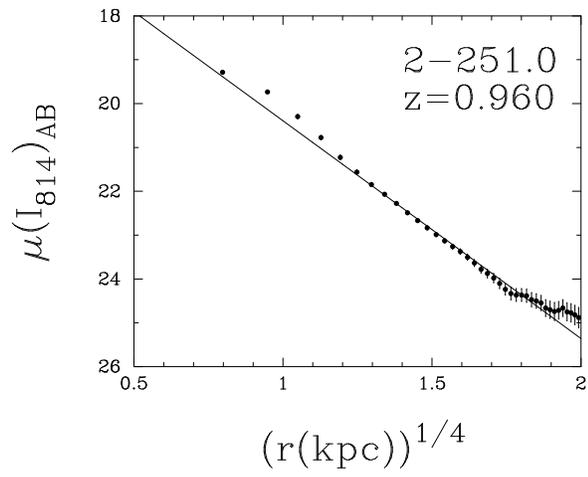
{Fig. 2 --- Azimuthally averaged radial surface brightness profiles of
our sample galaxies in the $I_{814}$-band.  (a) Red galaxies of which colors
are consistent with that of a passively evolving galaxy.  (b) Galaxies
of which colors are bluer than that of a passively evolving galaxy (see
also Figure 1).  Filled circles show the data points and solid lines
show the fitted lines with the $r^{1/4}$ law.}

\noindent 
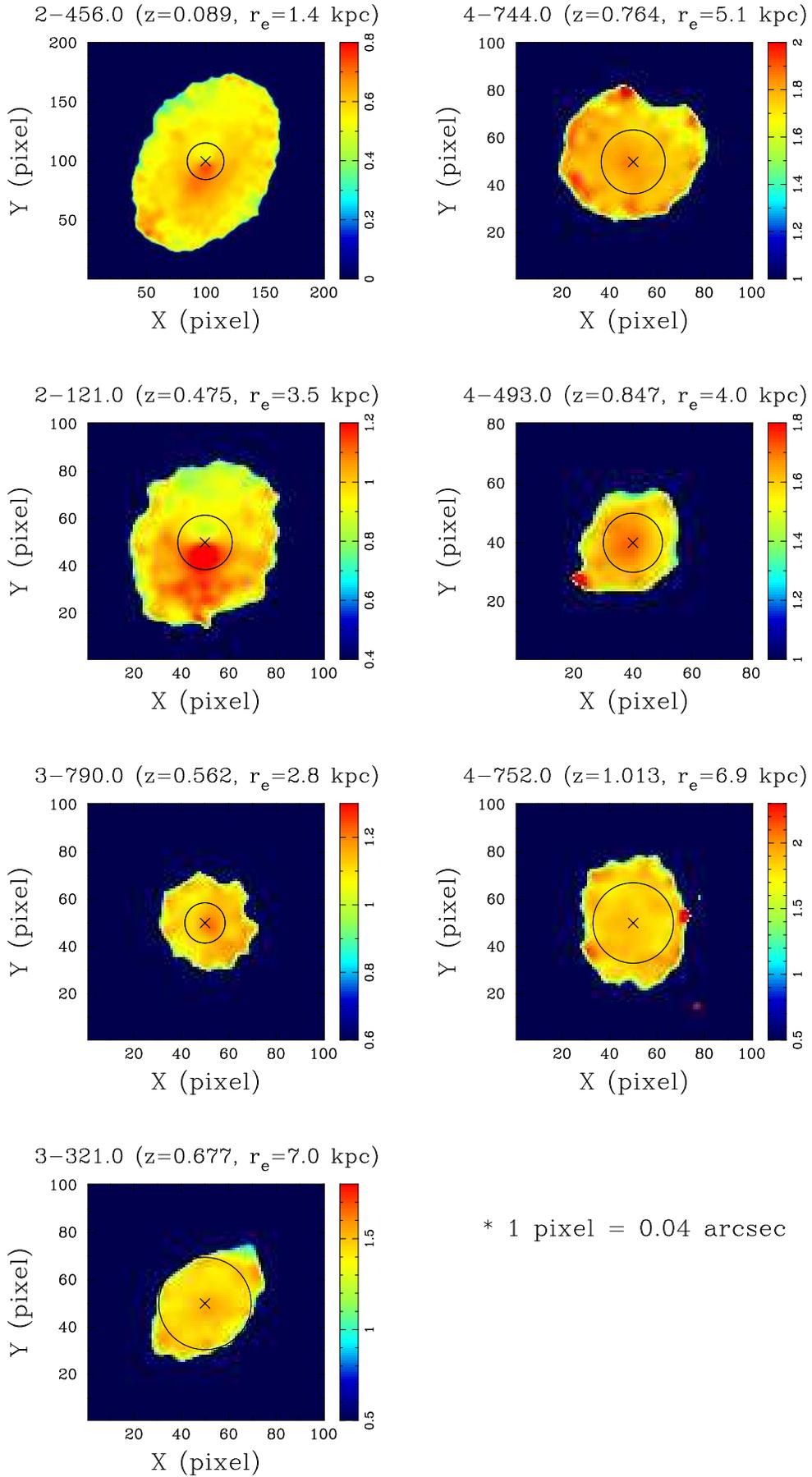
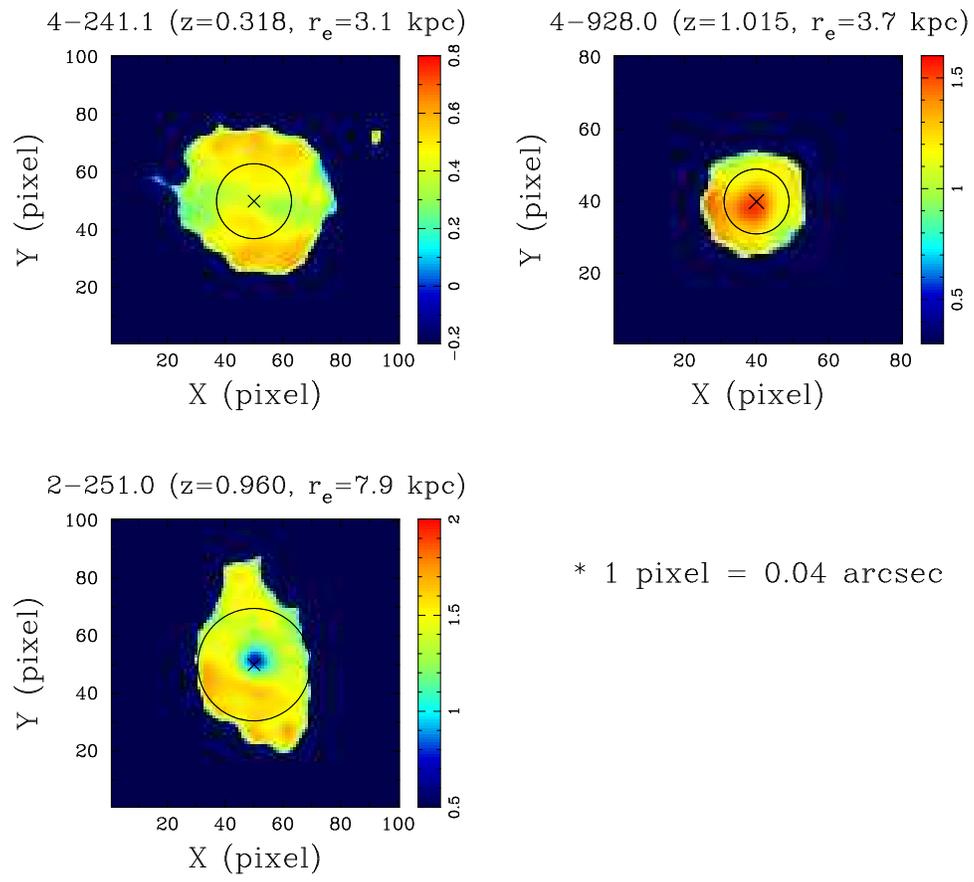
{Fig. 3 --- $V_{606} - I_{814}$ color maps for (a) the red galaxies whose
colors are consistent with that of a passively evolving galaxy, and (b)
the ``blue'' ellipticals.  Plus sign of each object denotes the
centroid of the galaxy in the $I_{814}$-band image.  The circle represents
an effective radius centered on the centroid.  Object ID, redshift, and
effective radius are also shown.  Note that a color bar is different
from galaxy to galaxy, to show the color distribution clearly.

\noindent 
{Fig. 4 --- Observed color gradients of the red elliptical galaxies
together with the model gradients by the age gradient.  Filled circles
represent the data points. Crosses refer to the data points which are
not used in deriving slopes of color profiles.  The models show the
predicted color gradients seen at each object's redshift. Solid lines
correspond to $\Delta(B - R)/\Delta \log (r/\rm r_{e}) = - 0.09 \pm
0.02$ mag/dex at $z = 0$, and the dotted lines correspond to $\Delta(B -
R)/\Delta \log (r/\rm r_{e}) = - 0.09 \pm 0.04$ mag/dex at $z = 0$.
Zero-points of the observed colors are slightly shifted by
$\Delta(V_{606} - I_{814})$ in Table 3 to compare the observed gradients
with the models.}

\noindent
{Fig. 5 --- Same as Fig. 4, but for the case of the metallicity gradient.}

\noindent
{Fig. 6 --- $V_{606} - I_{814}$ color gradient versus redshift for our
red sample galaxies.  Filled circles represent color gradients of the
sample ellipticals, which are determined within effective radii without
the innermost regions of the galaxies.  Solid and dotted lines indicate
the evolutions of color gradients caused by the metallicity gradient and
the age gradient, respectively.  (See text for the model.)}

\noindent
{Fig. 7 --- Color gradients of the galaxies which have relatively blue
colors. The color gradient of 4-928.0, which has a rather steep color
gradient, is plotted together with a model color
gradient made by age gradient fitted to the observed one by tuning ages
of stellar populations in the model galaxy. (See text for the model.)
The observed color for 4-928.0 is shifted by 0.4 mag to compare
the model gradients.}

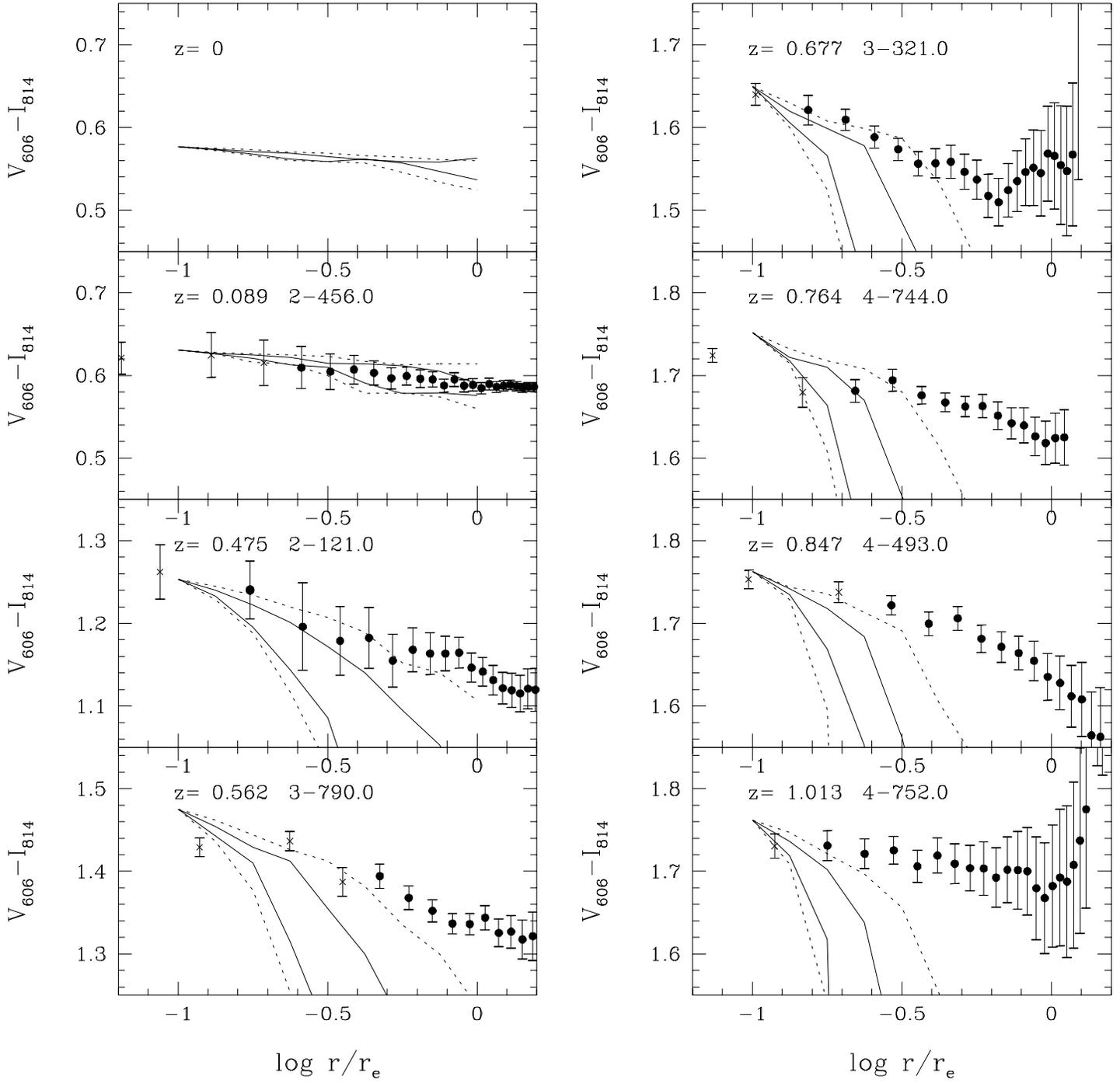
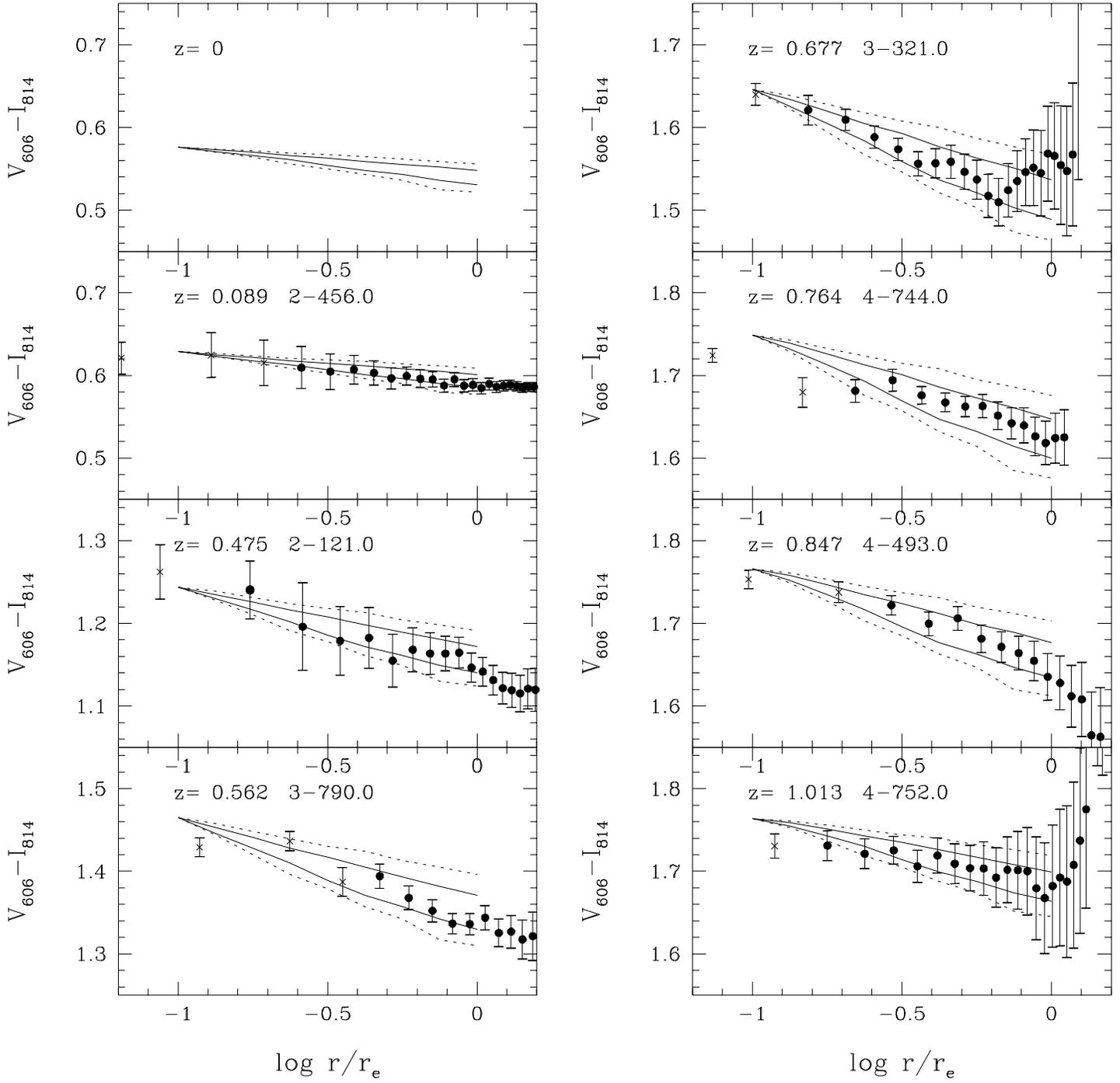
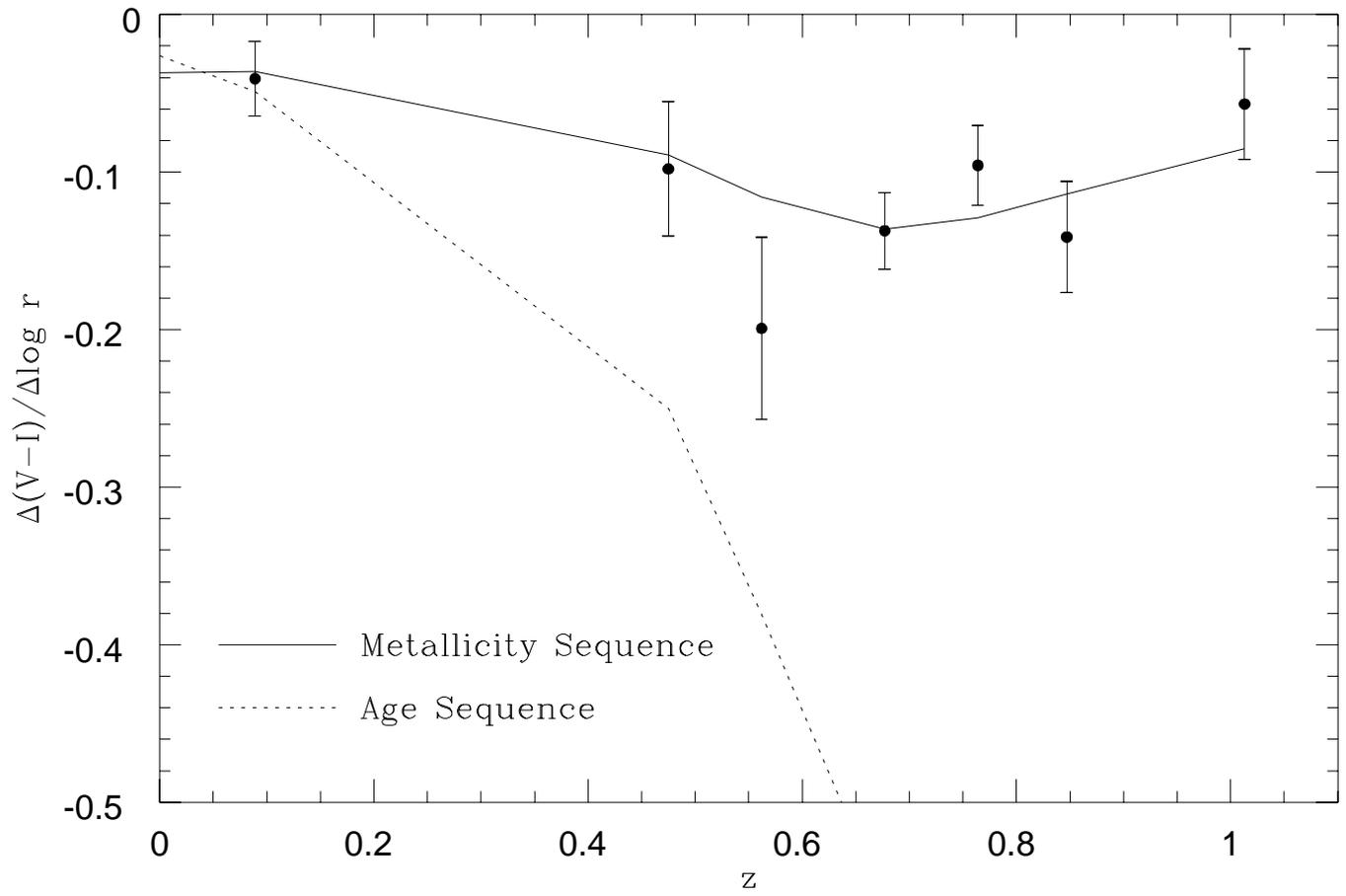
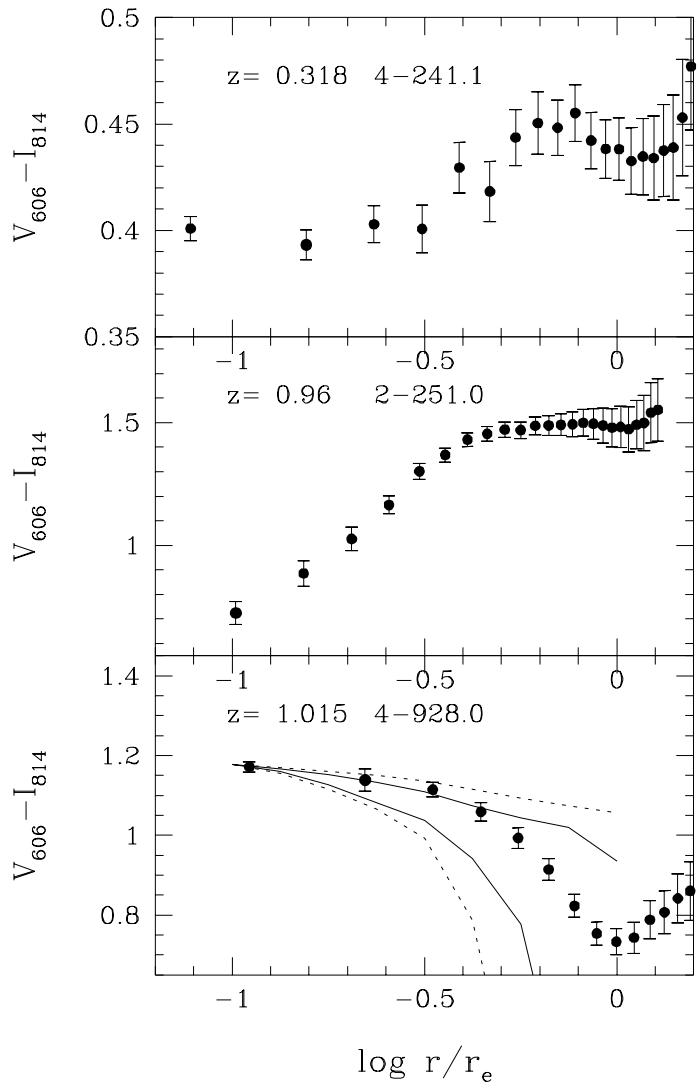
\end{document}